\documentclass[aps,prd,nofootinbib,amsmath,amssymb,superscriptaddress,onecolumn,11pt]{revtex4}
\usepackage{txfonts}
\usepackage{graphicx}
\usepackage{dcolumn}
\usepackage{bm}
\usepackage{amssymb}
\usepackage{latexsym}
\usepackage[colorlinks, linkcolor=blue, citecolor=blue, urlcolor=blue]{hyperref}

\newcommand{\be}{\begin{equation}}
\newcommand{\ee}{\end{equation}}
\newcommand{\bq}{\begin{eqnarray}}
\newcommand{\eq}{\end{eqnarray}}

\begin{document}

\title{Holographic dark energy in a universe with spatial curvature and massive neutrinos: a full Markov Chain Monte Carlo exploration}

\author{Yun-He Li}
\email{liyh19881206@126.com} \affiliation{Department of Physics,
College of Sciences, Northeastern University, Shenyang 110004,
China}

\author{Shuang Wang}
\email{swang@mail.ustc.edu.cn} \affiliation{Department of Physics,
College of Sciences, Northeastern University, Shenyang 110004,
China} \affiliation{Homer L. Dodge Department of Physics \&
Astronomy, Univ. of Oklahoma, 440 W Brooks St., Norman, OK 73019,
U.S.A.}

\author{Xiao-Dong Li}
\email{renzhe@mail.ustc.edu.cn} \affiliation{Institute of
Theoretical Physics, Chinese Academy of Sciences, Beijing 100190,
China} \affiliation{Kavli Institute for Theoretical Physics China,
Chinese Academy of Sciences, Beijing 100190, China}

\author{Xin Zhang\footnote{Corresponding author}}
\email{zhangxin@mail.neu.edu.cn} \affiliation{Department of Physics,
College of Sciences, Northeastern University, Shenyang 110004,
China} \affiliation{Center for High Energy Physics, Peking
University, Beijing 100080, China}

\begin{abstract}

In this paper, we report the results of constraining the holographic
dark energy model with spatial curvature and massive neutrinos,
based on a Markov Chain Monte Carlo global fit technique. The cosmic
observational data include the full WMAP 7-yr temperature and
polarization data, the type Ia supernova data from Union2.1 sample,
the baryon acoustic oscillation data from SDSS DR7 and WiggleZ Dark
Energy Survey, and the latest measurements of $H_0$ from HST. To
deal with the perturbations of dark energy, we adopt the
parameterized post-Friedmann method. We find that, for the simplest
holographic dark energy model without spatial curvature and massive
neutrinos, the phenomenological parameter $c<1$ at more than
$4\sigma$ confidence level. The inclusion of spatial curvature
enlarges the error bars and leads to $c<1$ only in about $2.5\sigma$
range; in contrast, the inclusion of massive neutrinos does not have
significant influence on $c$. We also find that, for the holographic
dark energy model with spatial curvature but without massive
neutrinos, the $3\sigma$ error bars of the current fractional
curvature density $\Omega_{k0}$ are still in order of $10^{-2}$; for
the model with massive neutrinos but without spatial curvature, the
$2\sigma$ upper bound of the total mass of neutrinos is $\sum
m_{\nu} < 0.48$ eV. Moreover, there exists clear degeneracy between
spatial curvature and massive neutrinos in the holographic dark
energy model, which enlarges the upper bound of $\sum m_{\nu}$ by
more than 2 times. In addition, we demonstrate that, making use of
the full WMAP data can give better constraints on the holographic
dark energy model, compared with the case using the WMAP ``distance
priors''.

\end{abstract}


\maketitle

\section{Introduction}

Observations of type Ia supernovae (SNIa) \cite{Riess}, cosmic
microwave background (CMB) \cite{spergel} and large scale structure
(LSS) \cite{Tegmark} all indicate that the universe is undergoing an
accelerating expansion. This implies the existence of a mysterious
component, called dark energy \cite{DEReview}, which has negative
pressure and takes the largest proportion of the total density in
the present universe. In the past fifteen years, lots of efforts
\cite{quint,phantom,k,tachyonic,hessence,Chaplygin,YMC,otherworks1,otherworks2}
have been made to understand dark energy, yet we still know little
about its nature.

In this paper we focus on the holographic dark energy model, which
is a quantum gravity approach to the dark energy problem
\cite{Holography}. In this model, the vacuum energy is viewed as
dark energy, and is related to the event horizon of the universe
when we require that the zero-point energy of the system should not
exceed the mass of a black hole with the same size \cite{cohen99}.
In this way, we have the holographic dark energy density \cite{li04}
\begin{equation}\label{eq:rhoHDE}
\rho_{de} = 3c^2M^2_{Pl}L^{-2},
\end{equation}
where $c$ is a dimensionless phenomenological parameter which plays
an important role in determining the properties of the holographic
dark energy, $M_{Pl}$ is the reduced Planck mass, and $L$ is the IR
cutoff length scale of the effective quantum field theory, which is
taken to be the future event horizon of the universe, $R_h$, in the
holographic model proposed by Li \cite{li04}, defined as
\begin{equation}\label{eq:rh}
R_h=a\int^\infty_t{dt \over a}=a\int^\infty_a{da\over Ha^2}.
\end{equation}
The holographic dark energy model has been proven to be a
competitive and promising dark energy candidate. It can
theoretically explain the coincidence problem \cite{li04}, and is
proven to be perturbational stable \cite{HDEstable}. Moreover, it is
favored by the observational data \cite{holoobs09}. For more studies
on the holographic dark energy model, see, e.g., Refs.
\cite{casimirHDE,refHDE,holode1,holode2,heal}.

It is fairly difficult to calculate the cosmological perturbations
of dark energy in the holographic dark energy model, because there
is a non-local effect making the calculation of perturbations
extremely hard to treat. Thus, in the past, for the CMB
observations, only the WMAP ``distance priors'' data were used to
constrain the holographic dark energy model in order to avoid the
inclusion of the perturbations of dark energy (see e.g. Refs.
\cite{CMBR1,CMBR2,decmb,CMBR3,ZL}). However, recently, it has been
realized that, in order to make progress one should first ignore the
non-local effect in the holographic dark energy model and directly
calculate the perturbations of holographic dark energy as if it is a
usual dynamical dark energy. Alone this line, in a recent work
\cite{Xu2012}, a global fitting analysis on the holographic dark
energy model was performed. 
However, in Ref.~\cite{Xu2012} the treatment of the gravity
instability problem concerning the phantom divide crossing is
absent.

In this paper, we shall perform a global fit analysis on the
holographic dark energy model with spatial curvature and massive
neutrinos. As argued in Ref. \cite{Clarkson:2007bc}, the numerical
studies of dynamical dark energy should include the spatial
curvature $\Omega_{k0}$ as a free parameter to be fitted alongside
the equation-of-state parameter (EOS) $w$ of dark energy. In
addition, the total mass of neutrinos $\sum m_{\nu}$ is also tightly
correlated with $w$ \cite{LiZhang:2012}. So, in our work, based on a
Markov Chain Monte Carlo (MCMC) global fit technique, we will
consider spatial curvature and massive neutrinos in the holographic
dark energy model, and will deeply analyze the influences of these
two factors on the fitting results. For a dynamical dark energy
model, one must be careful about the treatment of perturbations in
dark energy when $w$ crosses $-1$. In this work, following the WMAP
team, we use the ``parameterized post-Friedmann'' (PPF) approach
\cite{PPF1} implemented in the CAMB code.

This paper is organized as follows. In Section \ref{sec:2}, we
derive the basic equations for the holographic dark energy in a
universe with spatial curvature and massive neutrinos. In Section
\ref{sec:3}, we discuss the holographic dark energy in a background
universe and also in a perturbed universe, and then we make a global
fit analysis on the models. At last, some concluding remarks are
given in Section \ref{sec:4}. In this work, we assume today's scale
factor $a_{0}=1$, so the redshift $z$ satisfies $z=a^{-1}-1$; the
subscript ``0'' always indicates the present value of the
corresponding quantity, and the unit with $c=\hbar=1$ is used.

\section{Holographic dark energy model with spatial curvature and massive
neutrinos}\label{sec:2}

In this section, we derive the basic equations for the holographic
dark energy model with spatial curvature and massive neutrinos. Then we briefly introduce the PPF description of dark energy perturbations.

\subsection{Background evolution of holographic dark energy in a non-flat universe}

In a spatially non-flat Friedmann-Robertson-Walker universe, the
Friedmann equation can be written as
\begin{equation}
\label{eq:FE1} 3M_{Pl}^2 H^2=\rho_k+\rho_{dm}+\rho_{de}+\rho_b+\rho_{\nu}+\rho_{r},
\end{equation}
where $\rho_k=-3M_{Pl}^2 K/a^2$ is the effective energy
density of the curvature component, $\rho_{dm}$, $\rho_{de}$,
$\rho_{b}$, $\rho_{\nu}$ and $\rho_{r}$ represent the energy dinsity
of dark matter, dark energy, baryon, massive neutrinos and
radiation, respectively. Notice that we adopt the approximate method
used in the five-year analysis of WMAP \cite{WMAP5}: dividing
neutrino component into the relativistic neutrinos and the massive
neutrinos, and including the relativistic neutrinos into radiation
component. For convenience, we define the fractional energy
densities of the various components, $\Omega_i=\rho_i/(3M_{Pl}^2H^2)$.
It is clear that
\begin{equation}
\label{eq:AllOmega} \Omega_k+\Omega_{dm}+\Omega_{de}+\Omega_b+\Omega_{\nu}+\Omega_{r}=1.
\end{equation}

In addition, the energy conservation equations for the various
components are in the form
\begin{equation}
\label{eq:EC}\dot\rho_i+3H(1+w_i)\rho_i=0,
\end{equation}
where $w_{1,2,3} = 0$ for dark matter, baryons and massive
neutrinos, $w_4 = 1/3$ for relativistic components, $w_5 = -1/3$ for
curvature and $w_6 = p_{de}/\rho_{de}$ for dark energy. Note that,
in this paper, a over dot always denotes the derivative with respect
to the cosmic time $t$. Combining Eqs.~(\ref{eq:AllOmega}) and
(\ref{eq:EC}) gives
\begin{equation}
 p_{de}=-\frac{2}{3}\frac{\dot H}{H^2}\rho_c-\rho_c-{1\over3}\rho_{r}+{1\over3}\rho_k,\label{eq:pde}
\end{equation}
which together with energy conservation equation (\ref{eq:EC}) for
dark energy gives
\begin{equation}
 \label{eq:OH2} 2(\Omega_{de}-1){\dot H\over
 H}+\dot\Omega_{de}+H(3\Omega_{de}-3+\Omega_k-\Omega_{r})=0.
\end{equation}


In a non-flat universe, the IR cut-off length scale $L$ takes the
form
\begin{equation}
\label{eq:L1} L=ar(t),
\end{equation}
where
$$r(t)={1\over\sqrt{|K|}}\textrm{sinn}
\Big(\sqrt{|K|}\int_t^{+\infty} {dt \over
a}\Big)={1\over\sqrt{|K|}}\textrm{sinn}
\Big(\sqrt{|K|}\int_{a(t)}^{+\infty} {da \over {Ha^2}}\Big),$$ with
$\textrm{sinn}(x)=\sin(x)$, $x$, and $\sinh(x)$ for $K>0$, $K=0$,
and $K<0$, respectively. From the energy density of the holographic
dark energy, we have
\begin{equation}
\label{eq:L0} \Omega_{de}={c^2 \over H^2L^2}.
\end{equation}
Substituting Eq.~(\ref{eq:L1}) into Eq.~(\ref{eq:L0}) and taking
derivative of Eq.~(\ref{eq:L0}) with respect to $t$,
one can get
\begin{equation}
\label{eq:OL1.3}
{\dot\Omega_{de}\over2\Omega_{de}}+H+{\dot H\over H}=\sqrt{{\Omega_{de}H^2\over c^2}-{K\over a^2}}.
\end{equation}


Combining Eq.~(\ref{eq:OH2}) with Eq.~(\ref{eq:OL1.3}), we finally
obtain the following two equations governing the dynamical evolution
of the holographic dark energy in a universe with spatial curvature
and massive neutrinos,
\begin{equation}
\label{eq:OH3}{1\over E(z)}{dE(z) \over dz}
=-{\Omega_{de}\over
1+z}\left({\Omega_k-\Omega_{r}-3\over2\Omega_{de}}+{1\over2}+\sqrt{{\Omega_{de}\over
c^2}+\Omega_k} \right),
\end{equation}
\begin{equation}
\label{eq:OH4}
{d\Omega_{de}\over dz}=
-{2\Omega_{de}(1-\Omega_{de})\over 1+z}\left(\sqrt{{\Omega_{de}\over
c^2}+\Omega_k}+{1\over2}-{\Omega_k-\Omega_{r}\over 2(1-\Omega_{de})}\right),
\end{equation}
where $E(z)\equiv H(z)/H_0$ is the dimensionless Hubble expansion rate, and
\begin{equation}
\Omega_{k}(z)=\frac{\Omega_{k0}(1+z)^2}{E(z)^2},\ \
\Omega_{r}(z)=\frac{\Omega_{r0}(1+z)^4}{E(z)^2}.
\end{equation}
The initial conditions are $E(0)=1$ and
$\Omega_{de}(0)=1-\Omega_{k0}-\Omega_{dm0}-\Omega_{b0}-\Omega_{\nu0}-\Omega_{r0}$.
Note also that $\Omega_{\nu0}$ can be expressed as \cite{WMAP7}
\begin{equation}
\Omega_{\nu0}=\frac{\sum m_{\nu}}{94 h^2 {\rm eV}},
\end{equation}
where $h$ is the reduced Hubble constant, and $\sum m_{\nu}$ is the
sum of neutrino masses. In addition, the value of $\Omega_{r0}$ is
determined by the WMAP 7-yr observations \cite{WMAP7}
\begin{equation}
\Omega_{r0} = 2.469 \times 10^{-5} h^{-2} (1+0.2271N_{eff}),
\end{equation}
where $N_{eff}=3.04$ is the effective number of neutrino species.
Thus, Eqs. (\ref{eq:OH3}) and (\ref{eq:OH4}) can be solved
numerically, and will be used in the data analysis procedure.

In this paper, we shall consider the following four cases: (a) the
model of holographic dark energy without spatial curvature and
massive neutrinos ($\Omega_{k0}=0$ and $\sum m_{\nu}=0$), denoted as
HDE; (b) the model of holographic dark energy with spatial curvature
but without massive neutrinos ($\Omega_{k0}\neq 0$ but $\sum
m_{\nu}=0$), denoted as KHDE; (c) the model of holographic dark
energy with massive neutrinos but without spatial curvature ($\sum
m_{\nu} \neq 0$ but $\Omega_{k0}=0$), denoted as VHDE; (d) the model
of holographic dark energy with spatial curvature and massive
neutrinos ($\Omega_{k0}\neq 0$ and $\sum m_{\nu} \neq 0$), denoted
as KVHDE.

\subsection{PPF description for the perturbations of holographic dark energy}

In this work we calculate the linear metric and matter density
perturbations by using the formalism of Ma and Bertschinger
\cite{MaBert}. Also, we are very careful about the treatment of the
divergence problem \cite{Zhao:2005} for the dark energy
perturbations when $w$ crosses $-1$. Following the WMAP team, we
deal with this issue by using the PPF code \cite{PPF1}. This code
supports a time-dependent EOS $w$ that is allowed to cross $-1$
multiple times \cite{PPF2} for the dark energy perturbations.
Moreover, it has been widely used in the literature to deal with the
perturbations of dark energy (see e.g. Ref. \cite{SDSS3}).

Now, we shall first briefly review the PPF description of dark
energy perturbations \cite{PPF1,PPF2}. The perturbations of dark
energy can be described by the following four variables, density
fluctuation $\delta\rho_{de}$, velocity $v_{de}$, pressure
fluctuation $\delta p_{de}$, and anisotropic stress $\Pi_{de}$. The
evolution equations of $\delta\rho_{de}$ and $v_{de}$ are given by
corresponding continuity and Navier-Stokes equations, and $\Pi_{de}$
vanishes for linear perturbations. To complete the system, one needs
to specify the relationship between $\delta p_{de}$ and
$\delta\rho_{de}$, which defines a sound speed $c_s^2=\delta
p_{de}^{(rest)}/\delta\rho_{de}^{(rest)}$, where ``rest'' denotes
the dark energy rest frame. In an arbitrary gauge, one can obtain
\begin{equation}
 \delta
 p_{de}=c_s^2\delta\rho_{de}+3(1+w)\rho_{de}\left(c_s^2-\frac{\dot{p}_{de}}{\dot{\rho}_{de}}\right)\frac{v_{de}}{k_H},
\end{equation}
where $k_H = k/aH$ with $k$ the wave number in the Fourier space.
However, evidently, a gravity instability appears in the evolution
of perturbations because of the divergence of
$\dot{p}_{de}/\dot{\rho}_{de}$, when $w$ crosses the phantom divide
($w=-1$).

The PPF description replaces this condition on the pressure
perturbations with a direct relationship between the momentum
density of the dark energy and that of other components on large
scales, namely,
\begin{equation}\label{Eq:limll}
 \lim_{k_H \ll 1}{4\pi G \over H^2}(1+w)\rho_{de}{v_{de}-v_T\over k_H}=-{1 \over 3}c_K f_\zeta(a)k_H v_T,
\end{equation}
where $c_K=1-3K/k^2$, ``$T$" denotes all other components excluding
the dark energy, and $f_\zeta(a)$ is a function of time only. The
PPF description can make an exact match at large scales to any given
system with an arbitrary $w$ by solving the full equations at
$k_H\rightarrow0$ and inferring $f_\zeta(a)$ for the evolution of
all other finite $k$ modes. However, it is sufficient for most
purposes to simply take $f_\zeta=0$ \cite{PPF1}. Note that once
$v_{de}$ is determined, $\delta p_{de}$ follows by momentum
conservation with no singularities encountered as $w$ crosses the
phantom divide.

Besides, PPF description of dark energy gives the density and
momentum components of dark energy with a single joint dynamical
variable $\Gamma$. Here we directly give these two relationships in
the synchronous gauge,
\begin{eqnarray}
  \rho_{de}\delta_{de} &=& - 3\rho_{de}(1+w) \frac{v_{de}}{k_H} -\frac{c_K k_H^2 H^2}{4 \pi G} \Gamma, \\
  \rho_{de}(1+w)v_{de} &=& \rho_{de}(1+w)v_T - \frac{k_H H^2}{4 \pi G}\frac{1}{F}\left[ S - \Gamma - \frac{\dot{\Gamma}}{H} +
  f_{\zeta}\frac{4\pi G(\rho_T+P_T)(v_T + k \alpha )}{k_H H^2}\right],
\end{eqnarray}
where $$F = 1 + 3 {4 \pi G a^2 \over k^2 c_K} (\rho_T + p_T),$$ and
$$S= - {4\pi G \over  H^2} \left[ f_\zeta
(\rho_T+p_T)-\rho_{de}(1+w) \right] {(v_T + k \alpha ) \over k_H}.$$
Here $\delta_{de}\equiv\delta\rho_{de}/\rho_{de}$ and $\alpha \equiv
a(\dot{h} + 6\dot{\eta}) / 2k^2$ with $h$ and $\eta$ the metric
perturbations in the synchronous gauge.

At last, to assure that the dark energy becomes smooth relative to
the matter inside a transition scale $c_s k_H=1$ while exactly
conserving energy and momentum locally, one takes
\begin{equation}
(1 + c_\Gamma^2 k_H^2) \left[\frac{\dot{\Gamma}}{H} + \Gamma + c_\Gamma^2 k_H^2 \Gamma \right]= S, \label{eqn:gammaeom}
\end{equation}
where the free parameter $c_\Gamma$ gives the transition scale
between the large scales and small scales in terms of the Hubble
scale, whose value needs to be calibrated in practice. It is found
that $c_{\Gamma}=0.4c_s$ could match the evolution of scalar field
models \cite{PPF1}. Note that the PPF approach doesn't define a
sound speed $c_s$ for dark energy, but its value may indirectly
influence the perturbations of dark energy, since we adopt
$c_{\Gamma}=0.4c_s$. The possible influences of $c_s$ on the
amplitude of dark energy perturbations can be found in
Refs.~\cite{DeDeo:2003te,Bean:2003fb,Hannestad:2005ak,Xia:2007km,dePutter:2010vy,Ballesteros:2010ks,Ansari:2011wv}.
As long as $c_s$ is close to 1, the dark energy does not cluster
significantly on sub-horizon scales. Therefore, following the
treatment in the CAMB and CMBFAST codes, in our analysis we set
$c_s$ to be 1.

For the details about the PPF approach, we refer the reader to
Refs.~\cite{PPF1,PPF2}. Note that the CAMB Einstein-Boltzmann
package has been modified to include PPF \cite{PPF1} and a version
for dark energy has been made publically
available\footnote{\url{http://camb.info/ppf/} (PPF module for
CAMB).}.
In our calculations, the initial condition for the dynamical
variable $\Gamma$ of PPF is set to be zero and the adiabatic initial
conditions are taken for other variables.

\section{Global fit analysis}\label{sec:3}

In this section, we shall make a global fit analysis on the
holographic dark energy models considered above. We modify the MCMC
package ``CosmoMC'' \cite{COSMOMC} to do the numerical calculations.
Our most general parameter space vector is:
\begin{equation}
\label{parameter} {\bf P} \equiv (\omega_{b}, \omega_{dm}, \Theta,
\tau, c, \Omega_{k0}, \sum m_{\nu}, n_{s}, A_{s}),
\end{equation}
where $\omega_{b}\equiv\Omega_{b0}h^{2}$ and
$\omega_{dm}\equiv\Omega_{dm0}h^{2}$, $\Theta$ is the ratio
(multiplied by 100) of the sound horizon to the angular diameter
distance at decoupling, $\tau$ is the optical depth to
re-ionization, $c$ is the phenomenological parameter of the
holographic dark energy model, $A_s$ and $n_s$ are the amplitude and
the spectral index of the primordial scalar perturbation power
spectrum. For the pivot scale, we set $k_{s0}=0.002$Mpc$^{-1}$ to be
consistent with the WMAP team \cite{WMAP7}.

\begin{figure}
\centering
\includegraphics[width=8cm]{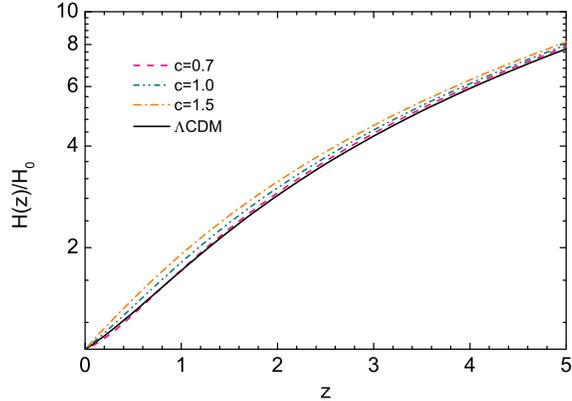}
\caption{The evolution of the dimensionless Hubble expansion rate
$H(z)/H_0$ in the holographic dark energy model. As an example, we
show the cases with $c=0.7$, $c=1.0$, and $c=1.5$. For a comparison,
the case in the $\Lambda$CDM model is also plotted. For the other
model parameters, we adopt their best-fit values given by the WMAP
7-yr observations.}\label{fig:Ez}
\end{figure}


In the computation of the CMB anisotropy, we include the WMAP 7-yr
temperature and polarization power spectra \cite{WMAP7} with the
routine for computing the likelihood supplied by the WMAP team
\cite{CMBcode}. For the SNIa, we make use of the recently released
580 SNIa data from the ``Union2.1'' sample \cite{Union2.1}, where
the systematic errors of SNIa are included in our analysis. For the
LSS information, we use the BAO data from the SDSS DR7
\cite{SDSSDR7} and WiggleZ Dark Energy Survey \cite{WiggleZ}. In
addition, we also use the latest Hubble space telescope (HST)
measurement of the Hubble constant, $H_{0} = 73.8 \pm
2.4$~km~s$^{-1}$~Mpc$^{-1}$ \cite{HSTWFC3}.

Before constraining the model parameter space, we study the effects
of the phenomenological parameter $c$ on the background evolution,
CMB power spectrum, and structure growth, by setting different $c$
and fixing the other model parameters. For simplicity, in these
examples we only consider the holographic dark energy model in a
flat universe.

First, let us have a look at the background evolution. The parameter
$c$ in the holographic dark energy model plays a very significant
role in determining the dynamical evolution of dark energy; for
details see Ref.~\cite{holoobs09}. In order to see the influence of
$c$ on the dynamics of expansion of the universe, we plot the
evolution of $H(z)/H_0$ for the holographic dark energy model in
Fig.~\ref{fig:Ez}. We take the values of $c$ to be 0.7, 1.0, and
1.5, as examples. We also plot the evolution of the $\Lambda$CDM
model for a comparison in this figure. For the other model
parameters, we adopt their best-fit values given by the WMAP 7-yr
observations~\cite{WMAP7}. From this figure, we can see that
different $c$ values in the holographic model lead to distinctly
different expansion histories. However, taking an appropriate value
for $c$, such as $c=0.7$, the holographic dark energy model will
produce a nearly indistinguishable background expansion history with
the $\Lambda$CDM model.

\begin{figure}
\centering
\includegraphics[width=8cm]{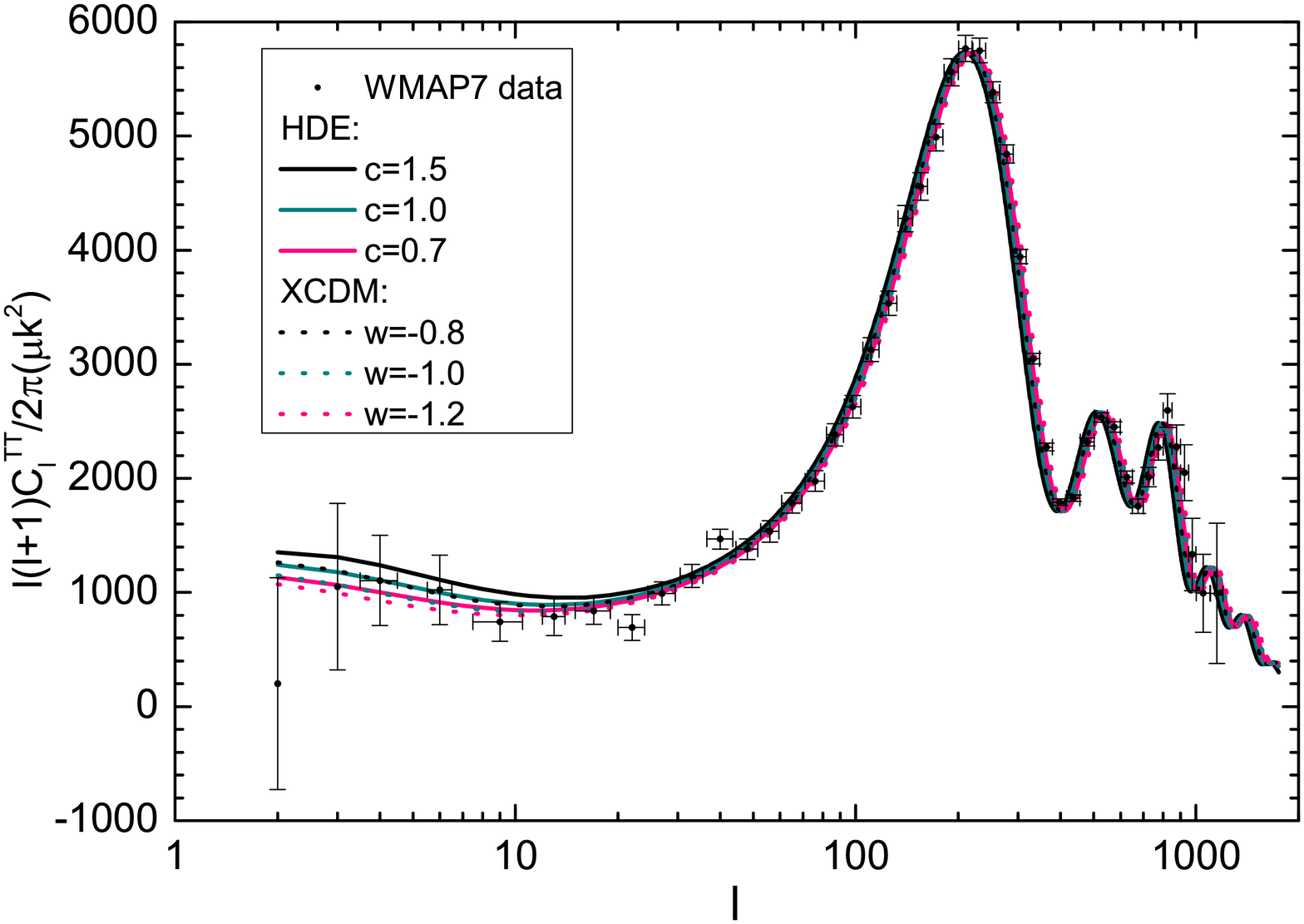}
\includegraphics[width=8cm]{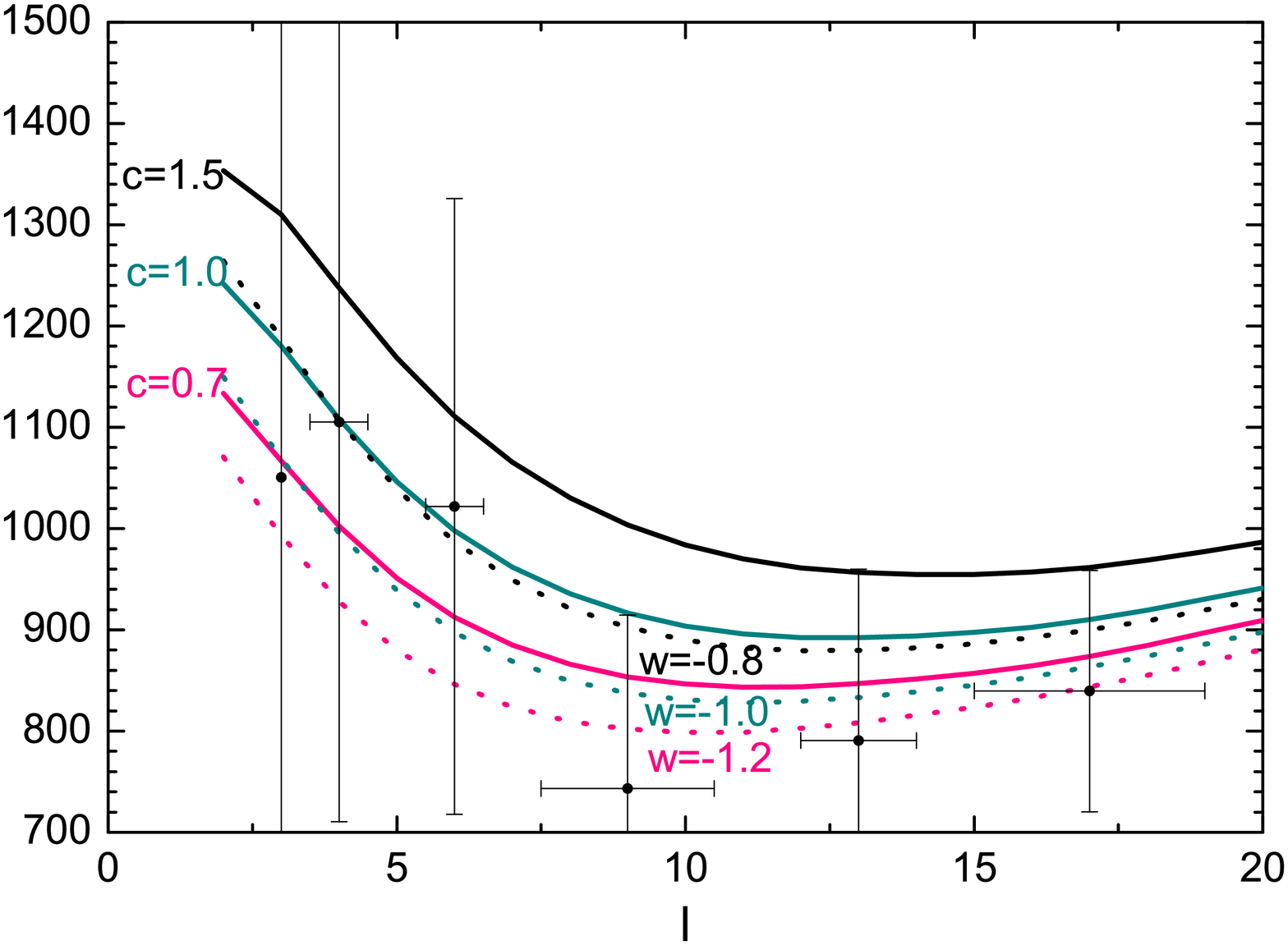}
\caption{The CMB $C_l^{TT}$ power spectrum in the holographic dark
energy model. The black dots with error bars denote the observed
data with their corresponding uncertainties from WMAP 7-yr results.
The solid lines denote the holographic dark energy models with
different $c$, and the dotted lines denote the XCDM models with
different $w$. For the other model parameters, we adopt their
best-fit values given by the WMAP 7-yr observations.}\label{cmbfig}
\end{figure}

Next, we want to see the situations in a perturbed universe. The
results of the CMB $C_l^{TT}$ power spectrum in the holographic dark
energy model are shown in Fig.~\ref{cmbfig}. For comparison, we also
plot the results of the XCDM models with different constant $w$. One
can see that, for the holographic models with different $c$, the
main difference appears at low ($l<20$) multipole momentum parts
which correspond to large scales. As seen in this figure, a smaller
$c$ will yield a smaller $C_l^{TT}$ at low $l$. Besides, since the
EOS of the holographic dark energy satisfies
$w=-\frac{1}{3}-\frac{2\sqrt{\Omega_{de}}}{3c}$ \cite{li04}, a
smaller $c$ will give a smaller $w$. So for the HDE model, a smaller
$w$ will lead to a smaller $C_l^{TT}$ at low $l$. Similarly, for the
XCDM model, a smaller $w$ will also lead to a smaller $C_l^{TT}$ at
low $l$. Thus, the results of the HDE and XCDM models are consistent
with each other.
We also notice that, for the CMB temperature angular power spectrum,
the holographic dark energy model with $c=0.7$ can also produce a
roughly same result with the $\Lambda$CDM model.

\begin{figure}
\centering
\includegraphics[width=8cm]{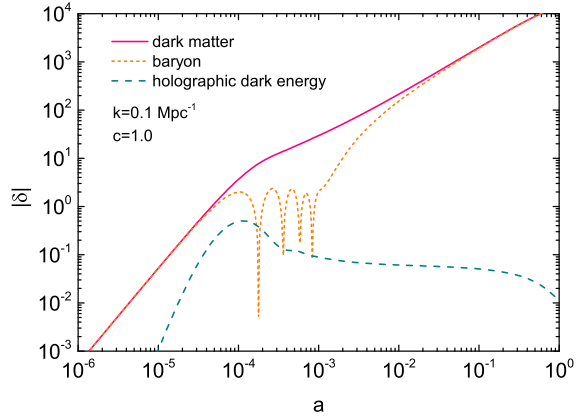}
\caption{Evolution of the density fields in the synchronous gauge
for $k=0.1$ Mpc$^{-1}$ in the holographic dark energy model with
$c=1$. The pink solid line, orange short-dashed line, and cyan
dashed line denote the density perturbations of dark matter, baryon,
and holograph dark energy, respectively. For the other model
parameters, we adopt their best-fit values given by the WMAP 7-yr
observations.}\label{fig:delta}
\end{figure}

\begin{figure}
\centering
\includegraphics[width=8cm]{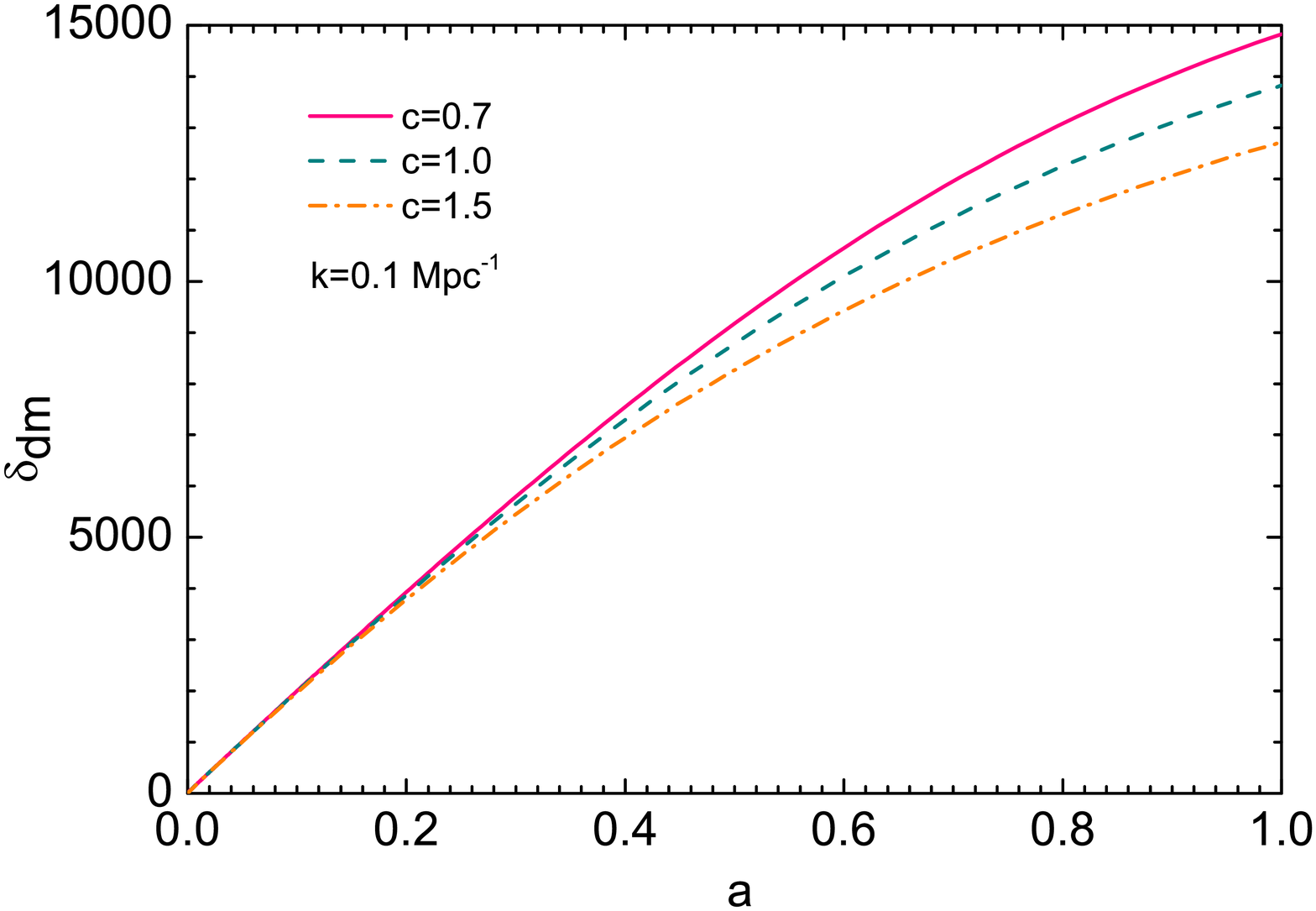}
\includegraphics[width=8cm]{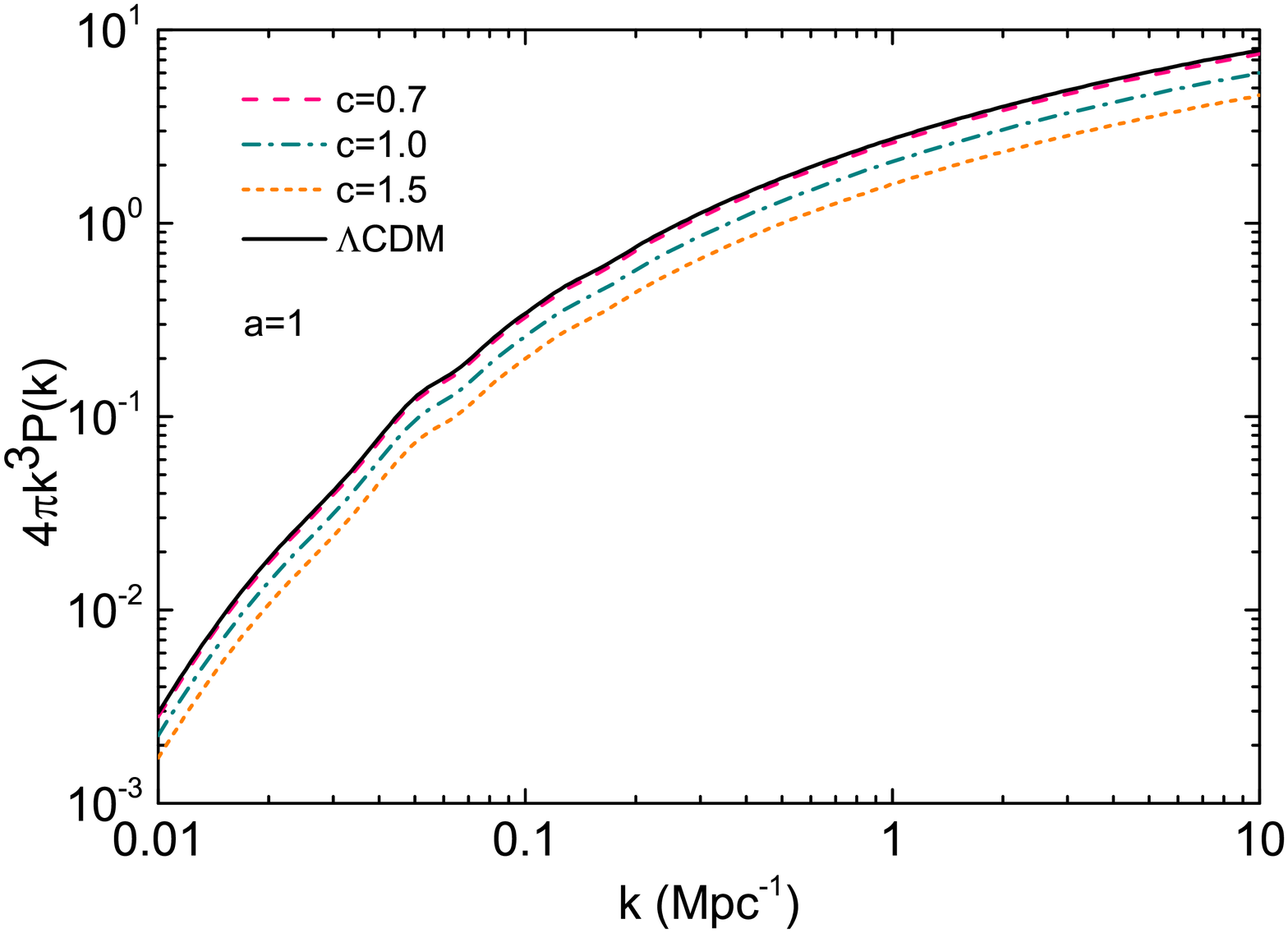}
\caption{Left panel: the evolution of the density perturbations of
cold dark matter for $k=0.1$ Mpc$^{-1}$. Right panel: the matter
power spectra at $z=0$. The cases of $c=0.7$, $c=1.0$, and $c=1.5$
in the holographic dark energy model are shown. For the other model
parameters, we adopt their best-fit values given by the WMAP 7-yr
observations.}\label{fig:deltac}
\end{figure}


Figure~\ref{fig:delta} shows the evolution of the density
perturbations for dark matter, baryon, and holographic dark energy,
in the HDE model with $c=1$ in the synchronous gauge. As a typical
example, we only plot the $k=0.1$ Mpc$^{-1}$ case. The behavior of
the density perturbations outside the horizon is strongly
gauge-dependent. In the synchronous gauge, all the $\delta$'s before
horizon crossing grow. After horizon crossing, the perturbations
come into causal contact and become nearly independent of the
coordinate choices. For the cold dark matter, after the horizon
crossing the fluctuations still monotonously grow. The mode with
$k=0.1$ Mpc$^{-1}$ enters the horizon before recombination, so the
baryons together with the photons oscillate acoustically while they
are coupled by Thomson scattering. The baryons decouple from the
photons at the recombination and then fall very quickly into the
potential wells formed around the cold dark matter, resulting in the
rapid growth of $\delta_b$, as seen in Fig.~\ref{fig:delta}. For the
holographic dark energy, we see clearly that after the horizon
crossing, the fluctuations in dark energy decrease quickly. So, the
dark energy does not cluster significantly on the sub-horizon size.
During the expansion of the universe, $\delta_{de}$ is always
smaller than $\delta_{dm}$ by several orders of magnitude, so the
dark energy perturbations almost do not affect the evolution of cold
dark matter perturbations. Hence, the impact of holographic dark
energy on the evolution of dark matter perturbations is almost only
from the background evolution. In the $\Lambda$CDM model, the
cosmological constant does not produce fluctuations, so its impact
on the evolution of dark matter density fluctuations is exactly from
the background evolution.

In order to know more about the large-scale structure in the
holographic dark energy model, we plot the evolution of dark matter
density perturbations and matter power spectrum with different $c$
values in Fig.~\ref{fig:deltac}. From the left panel of
Fig.~\ref{fig:deltac}, we can directly see the impact of the value
of $c$ on the growth of structure. Again, we take the case of
$k=0.1$ Mpc$^{-1}$ as an example. Once the accelerated expansion
begins, the growth of linear perturbations effectively tends to end,
since the Hubble damping time becomes shorter than the timescale for
perturbation growth. For larger $c$ and fixed present dark energy
density $\Omega_{de0}$, dark energy comes to dominate earlier,
causing the growth of linear perturbations to end earlier; this
explains the amplitude of perturbation with smaller $c$ is larger at
all redshifts until today. In the right panel of
Fig.~\ref{fig:deltac}, we show the matter power spectra with
different values of $c$ at $z=0$ in the holographic dark energy
model. We find that the impact of $c$ on the small scales is
slightly larger than that on the large scales. We also make a
comparison with the $\Lambda$CDM model. It is found that the HDE
model with $c=0.7$ and the $\Lambda$CDM model produce the almost
undistinguishable matter power spectra.

So far, we have deeply analyzed the holographic dark energy model in
the aspects of background universe, CMB, and LSS. We find that the
parameter $c$ plays an important role in all these aspects. We also
find that the HDE model with $c\simeq0.7$ strongly degenerates with
the $\Lambda$CDM model. In the following, we perform a global fit on
the holographic dark energy models by using the full CMB data
combined with the data of SNIa, BAO, and Hubble constant.

\begin{figure}
\centering
\includegraphics[width=13cm]{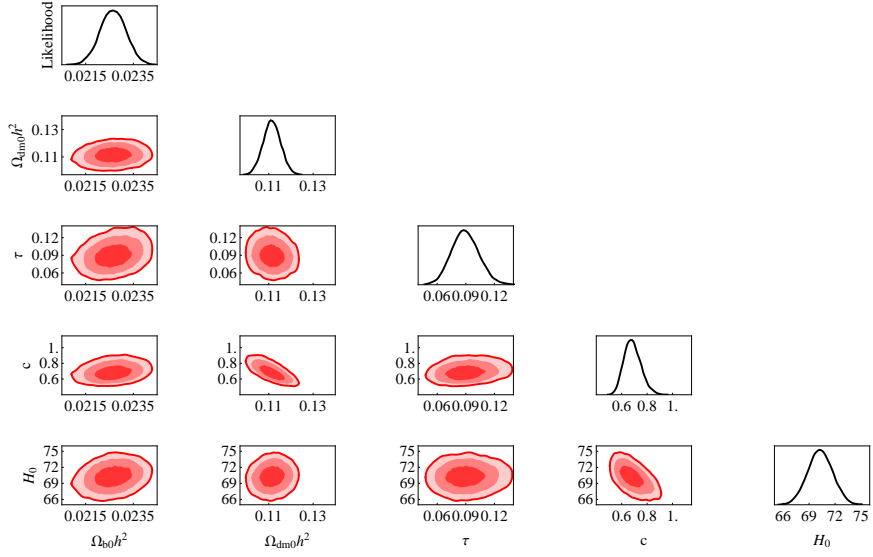}
\caption{The 1D marginalized distributions of individual parameters
and 2D marginalized 1--3$\sigma$ CL contours, for the HDE
model.}\label{fig:HDE}
\end{figure}

\begin{figure}
\centering
\includegraphics[width=8cm]{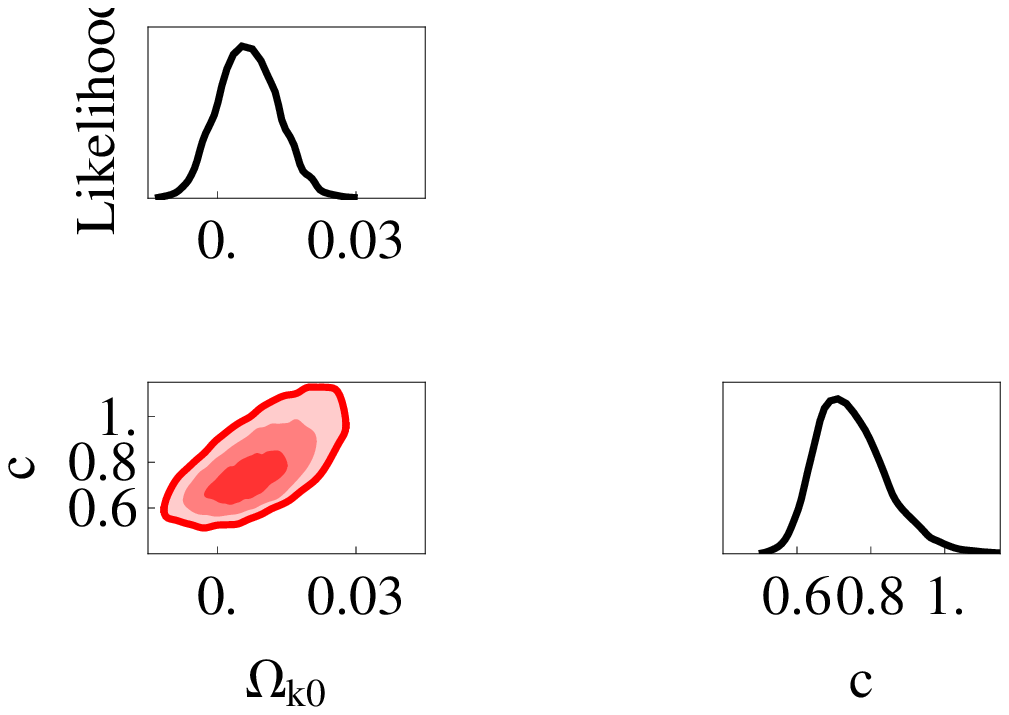}
\includegraphics[width=8cm]{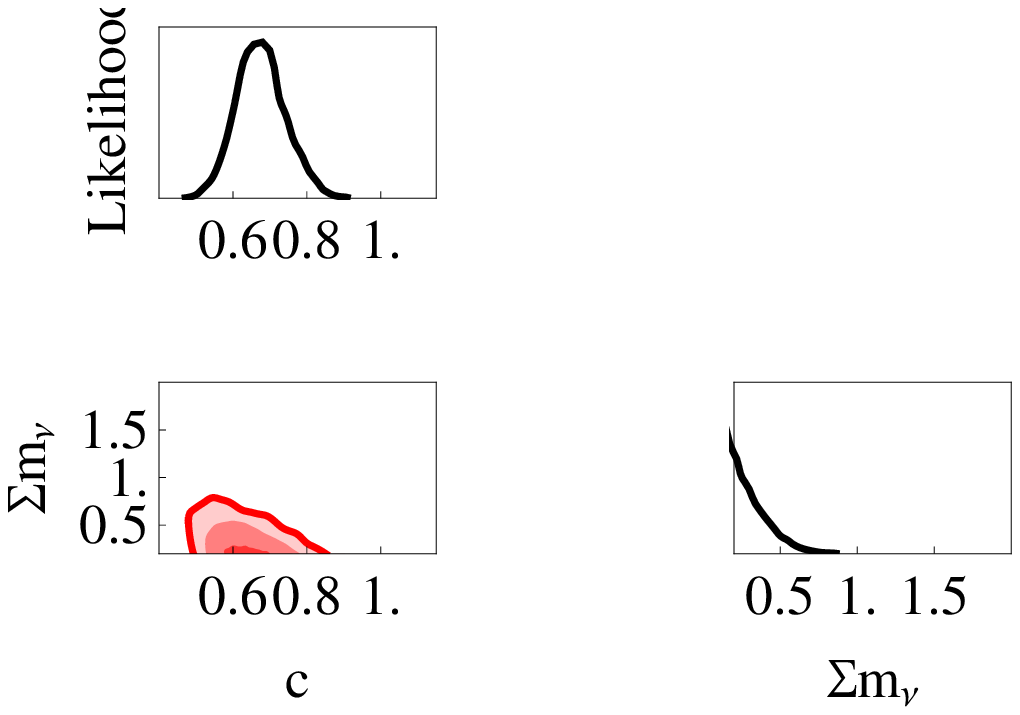}
\caption{The 1D marginalized distributions of individual parameters
and 2D marginalized 1--3$\sigma$ CL contours, for the KHDE model
(left panel) and the VHDE model (right panel). Here only the most
relevant parameters are presented.}\label{fig:KHDEVHDE}
\end{figure}

\begin{figure}
\centering
\includegraphics[width=17cm]{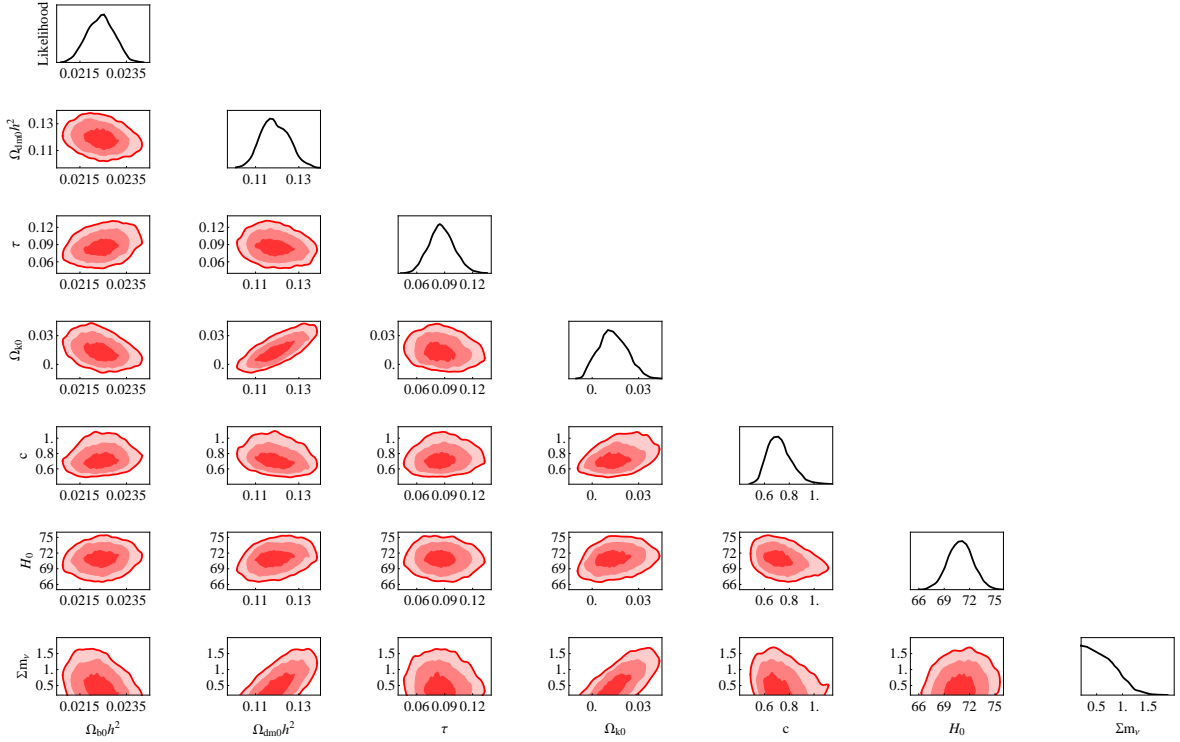}
\caption{The 1D marginalized distributions of individual parameters
and 2D marginalized 1--3$\sigma$ CL contours, for the KVHDE
model.}\label{fig:KVHDE}
\end{figure}

For the four holographic dark energy models considered, we run 16
independent chains with $\sim$ 20,000 samples in each chain. The
following priors to model parameters are adopted:
$\omega_{b}\in[0.005,0.1]$, $\omega_{dm}\in[0.01,0.99]$,
$\Theta\in[0.5,10]$, $\tau\in[0.01,0.8]$, $c\in[0.3,1.5]$,
$\Omega_{k0}\in[-1,1]$, $n_{s}\in[0.5,1.5]$, and
$\log[10^{10}A_{s}]\in[2.7, 4]$. Furthermore, the hard coded prior
on the cosmic age $10~\text{Gyr}<t_{0}<\text{20~Gyr}$ is also
imposed and the parameter $f_{\nu}$ (the definition is the ratio
between the current densities of massive neutrinos and cold dark
matter) ranges from 0 to 1. We did not use the option MPI Check
Limit Converge, but we find the chains converge at acceptable
values. Values of $-\ln \mathcal{L}_{max}$ and convergence (the
worst $e$-values [the variance(mean)/mean(variance) of 1/2 chains]
$R-1$) for the four models are listed in the Table I. Table
\ref{table1} summarizes the fitting results, including the best-fit
and 1--3$\sigma$ values of the relevant cosmological parameters, as
well as the $2\sigma$ upper bounds of $\sum m_{\nu}$, for the
considered models. Moreover, we also list the maximal confidence
levels for $c<1$, in this table. In addition, in
Figs.~\ref{fig:HDE}--\ref{fig:KVHDE}, we plot the 1D marginalized
distributions of individual parameters,\footnote{We find that the 1D
marginalized posteriors and mean likelihoods of $c$, $\sum m_{\nu}$,
and $\Omega_{k0}$ are close to each other and have the same shapes,
thus here we only show the marginalized posteriors.} as well as the
2D marginalized 1--3$\sigma$ CL contours, for these models. Let us
discuss them in detail in what follows.

\begin{table*}\caption{Fitting results of the holographic dark energy
models.}\label{table1}
\begin{center}
\begin{tabular}{cc   cc    cc   cc   cc}
\hline\hline Parameters & & HDE  &  & KHDE & & VHDE & & KVHDE&
\\ \hline
$\Omega_{dm0}h^2$
                   &&$0.110^{+0.005+0.009+0.013}_{-0.002-0.006-0.010}$&
                   &$0.113^{+0.007+0.012+0.016}_{-0.002-0.007-0.012}$&
                   &$0.110^{+0.006+0.009+0.012}_{-0.002-0.005-0.009}$&
                   &$0.117^{+0.009+0.015+0.021}_{-0.004-0.009-0.016}$&\\
$100\Omega_{b0}h^2$
                   &&$2.259^{+0.061+0.115+0.159}_{-0.048-0.100-0.152}$&
                   &$2.246^{+0.063+0.121+0.159}_{-0.045-0.098-0.142}$&
                   &$2.272^{+0.044+0.096+0.144}_{-0.069-0.105-0.155}$&
                   &$2.256^{+0.028+0.097+0.146}_{-0.066-0.128-0.173}$&\\
$c$
                   &&$0.680^{+0.064+0.135+0.222}_{-0.066-0.119-0.159}$&
                   &$0.702^{+0.104+0.232+0.393}_{-0.063-0.102-0.176}$&
                   &$0.708^{+0.014+0.111+0.159}_{-0.099-0.153-0.215}$&
                   &$0.733^{+0.037+0.185+0.321}_{-0.107-0.170-0.230}$&\\
$\Omega_{k0}$      &&N/A &
                   &$0.004^{+0.009+0.016+0.023}_{-0.004-0.010-0.015}$&
                   &N/A&
                   &$0.010^{+0.010+0.020+0.032}_{-0.004-0.014-0.018}$&\\
$\sum m_{\nu}$ (eV)
                   &&N/A&
                   &N/A&
                   &$\sum m_{\nu}<0.48$ (2$\sigma$)&
                   &$\sum m_{\nu}<1.17$ (2$\sigma$)&\\
$H_0$ (km/s/Mpc)             &&$70.3^{+1.3+2.9+4.2}_{-1.3-2.8-4.0}$&
                   &$70.7^{+1.7+3.1+4.6}_{-1.2-2.6-4.2}$&
                   &$69.6^{+1.8+3.3+4.8}_{-0.9-2.1-3.5}$&
                   &$71.0^{+1.1+2.8+4.2}_{-1.2-2.8-4.1}$&\\
$\tau$
                   &&$0.087^{+0.015+0.030+0.047}_{-0.013-0.026-0.038}$&
                   &$0.087^{+0.014+0.029+0.046}_{-0.015-0.026-0.037}$&
                   &$0.087^{+0.015+0.031+0.047}_{-0.011-0.023-0.036}$&
                   &$0.085^{+0.012+0.029+0.044}_{-0.009-0.024-0.035}$&\\
$\Theta$
                   &&$1.039^{+0.003+0.006+0.008}_{-0.002-0.004-0.006}$&
                   &$1.039^{+0.003+0.005+0.008}_{-0.001-0.004-0.006}$&
                   &$1.040^{+0.002+0.005+0.007}_{-0.003-0.005-0.007}$&
                   &$1.040^{+0.001+0.004+0.007}_{-0.002-0.005-0.008}$&\\
$n_s$
                   &&$0.966^{+0.018+0.029+0.041}_{-0.010-0.020-0.031}$&
                   &$0.968^{+0.009+0.024+0.035}_{-0.015-0.025-0.037}$&
                   &$0.969^{+0.012+0.025+0.039}_{-0.011-0.023-0.032}$&
                   &$0.969^{+0.008+0.020+0.034}_{-0.016-0.032-0.046}$&\\
$\log[10^{10}A_s]$
                   &&$3.181^{+0.036+0.070+0.107}_{-0.037-0.078-0.112}$&
                   &$3.188^{+0.039+0.081+0.120}_{-0.026-0.071-0.108}$&
                   &$3.172^{+0.039+0.072+0.115}_{-0.029-0.068-0.106}$&
                   &$3.191^{+0.039+0.097+0.123}_{-0.022-0.065-0.118}$&\\
$\Omega_{de0}$
                   &&$0.731^{+0.009+0.022+0.032}_{-0.015-0.029-0.043}$&
                   &$0.725^{+0.012+0.023+0.036}_{-0.018-0.036-0.053}$&
                   &$0.726^{+0.015+0.026+0.035}_{-0.012-0.025-0.038}$&
                   &$0.714^{+0.011+0.030+0.042}_{-0.023-0.048-0.073}$&\\
Age (Gyr)
                   &&$13.875^{+0.076+0.183+0.284}_{-0.123-0.230-0.328}$&
                   &$13.708^{+0.115+0.410+0.671}_{-0.422-0.594-0.870}$&
                   &$13.869^{+0.093+0.241+0.378}_{-0.129-0.209-0.289}$&
                   &$13.480^{+0.244+0.566+0.819}_{-0.176-0.507-0.804}$&\\
$\Omega_{m0}$
                   &&$0.269^{+0.015+0.029+0.043}_{-0.009-0.022-0.032}$&
                   &$0.271^{+0.014+0.029+0.045}_{-0.009-0.023-0.032}$&
                   &$0.274^{+0.012+0.025+0.038}_{-0.015-0.026-0.035}$&
                   &$0.276^{+0.015+0.031+0.049}_{-0.010-0.023-0.033}$&\\
$z_{re}$
                   &&$10.580^{+1.184+2.312+3.552}_{-1.106-2.251-3.447}$&
                   &$10.607^{+1.128+2.275+3.533}_{-1.187-2.317-3.472}$&
                   &$10.504^{+1.235+2.439+3.674}_{-0.908-2.075-3.210}$&
                   &$10.526^{+1.143+2.532+3.455}_{-0.703-2.083-3.213}$&\\
\hline CL for $c<1$  &&4.2$\sigma$ & & 2.5$\sigma$  & & 4.6$\sigma$  & &2.7$\sigma$&\\
\hline $-\ln \mathcal{L}_{max}$ / convergence  &&4009.2/0.0085 & & 4008.9/0.0202  & & 4009.2/0.0384  & &4008.7/0.0434&\\
\hline\hline
\end{tabular}
\label{table2}
\end{center}
\end{table*}

The fitting results of the HDE model are shown in
Fig.~\ref{fig:HDE}. We find that the values of $c$ in 1--3$\sigma$
regions are $c = 0.680^{+0.064+0.135+0.222}_{-0.066-0.119-0.159}$;
note that the error bars are slightly smaller than those obtained by
using the WMAP ``distance priors'' \cite{ZL}. Moreover, we find that
$c < 1$ at $4.2\sigma$ CL, which means that the future universe will
be dominated by phantom energy and will end up with a ``Big Rip''
(cosmic doomsday) at more than $4\sigma$ CL. In Ref.~\cite{heal} it
has been demonstrated that $c<1$ may lead to the ruin of the
theoretical foundation---the effective quantum field theory---of the
holographic dark energy scenario. To rescue the holographic scenario
of dark energy, one may employ the braneworld cosmology and
incorporate the extra-dimension effects into the holographic theory
of dark energy. It has been found \cite{heal} that such a mend could
erase the big-rip singularity and leads to a de Sitter finale for
the holographic cosmos. In addition, if there is some direct,
non-gravitational interaction between holographic dark energy and
dark matter, and such an interaction satisfies some conditions, the
big rip can also be avoided \cite{ZL,KIHDE}.

Figure~\ref{fig:KHDEVHDE} shows the results of the KHDE and VHDE
models. In this figure, we only present the most interesting
parameters for our discussions. It is seen that, compared with the
HDE model, the KHDE model slightly favors a larger best-fit value of
$c$. Moreover, the error bars of $c$ are also enlarged. As shown in
Fig.~\ref{fig:HDE}, for the HDE model, we have $c<1$ at more than
$4\sigma$ CL. However, as shown in the left panel of
Fig.~\ref{fig:KHDEVHDE}, after considering spatial curvature, we
have $c<1$ only in $2.5\sigma$ range.

This figure also shows the degeneracy situation of $\Omega_{k0}$ and
$c$, in the KHDE model. It is clear that $\Omega_{k0}$ and $c$ are
in positive correlation. This result is well consistent with our
pervious work (see Fig. 4 of Ref.~\cite{KIHDE}; notice that the
convention of $\Omega_{k0}$ in this paper is different from that in
Ref.~\cite{KIHDE}). The best-fit value for $\Omega_{k0}$ is very
close to zero, i.e., $\Omega_{k0}=0.004$. The 2$\sigma$ range of the
spatial curvature is $-0.006<\Omega_{k0}<0.020$. Actually, the
$3\sigma$ error bars of $\Omega_{k0}$ are still fairly small, also
in order of $10^{-2}$.

The constraints on the VHDE model are shown in the right panel of
Fig.~\ref{fig:KHDEVHDE}. One can see that, compared with the HDE
model, the VHDE model also slightly favors a larger best-fit value
of $c$. However, for the VHDE model, the changes on the error bars
of $c$ are quite different from the KHDE model: in the KHDE model,
both the upper and the lower bounds of $c$ are enlarged comparing to
the HDE model; while in the VHDE model, although the lower bounds of
$c$ are enlarged, the upper bounds of $c$ are reduced comparing to
the HDE model. Moreover, for the VHDE model, we have $c<1$ at
$4.6\sigma$ CL, which is quite similar to the result of the simplest
HDE model. Therefore, we can conclude that the inclusion of massive
neutrinos does not have significant influence on the
phenomenological parameter $c$.

From the right panel of Fig.~\ref{fig:KHDEVHDE}, one can also see that
there is only weak negative correlation between $\sum m_{\nu}$ and $c$,
in the VHDE model. In addition, we obtain the upper bound of
the total mass of neutrinos in the holographic dark energy model,
$\sum m_{\nu} < 0.48$ eV at $2\sigma$ CL. This is the first result
of the neutrino mass in the holographic dark energy model. For
comparison, we mention here some results of neutrino mass in other
dark energy scenarios. For a flat $\Lambda$CDM model, i.e., $w=-1$
and $\Omega_{k0}=0$, Komatsu et al. \cite{WMAP7} found that the
WMAP+BAO+$H_0$ limit is $\sum m_{\nu}<0.58$ eV ($95\%$ CL). For a
constant $w$ model the results given by Komatsu et al. \cite{WMAP7}
are: $\sum m_{\nu}<0.71$ eV ($95\%$ CL) from WMAP+LRG+$H_0$, and
$\sum m_{\nu}<0.91$ eV ($95\%$ CL) from WMAP+BAO+SN (where SN is the
Constitution sample).


The results for the most sophisticated case, i.e. the KVHDE model,
are shown in Fig.~\ref{fig:KVHDE}. We find that, for the KVHDE
model, we have $c<1$ only at the $2.7\sigma$ level.

This figure also shows the degeneracy situations of various
parameters, in the KVHDE models. In the KHDE model we find that
$\Omega_{k0}$ and $c$ are in positive correlation, and in the VHDE
model we do not observe significant correlation between $\sum
m_{\nu}$ and $c$. So, we believe that the parameters $\Omega_{k0}$
and $\sum m_{\nu}$ should be in positive correlation. Indeed, in the
KVHDE model, we find that the result is in accordance with the
expectation. In addition, when simultaneously considering spatial
curvature and massive neutrinos in the holographic dark energy
model, the parameter space of $(\Omega_{k0}, \sum m_{\nu})$ is
greatly amplified. For example, the upper bound of $\sum m_{\nu}$ is
enlarged by more than 2 times comparing to the VHDE model.

An interesting phenomenon in Table \ref{table1} is that the
marginalized ranges of some cosmological parameters narrow when we
add additional parameters. For example, compared with the HDE and
KHDE models, the upper error bars of $c$ shrink when we add $\sum
m_\nu$ in the VHDE and KVHDE models. This phenomenon is
understandable. In Fig. \ref{fig:KVHDE}, we saw that a larger
neutrino mass favors a smaller $c$, thus adding $\sum m_{\nu}$ into
the fit will drag $c$ to smaller values and shrink its upper error
bars. Similar phenomenon also appears in other parameters. Due to
the negative correlation with $\sum {m_\nu}$, we find narrow upper
error bars of $100\Omega_{b0}h^2$ in the VHDE and KVHDE models, and
the narrow upper error bars of $n_s$ in KHDE and KVHDE models are
due to its negative correlation with $\Omega_{k0}$.\footnote{Since
$n_s$ is not our major focus, its contours are not plotted.} 

\begin{figure}
\centering
\includegraphics[width=8cm]{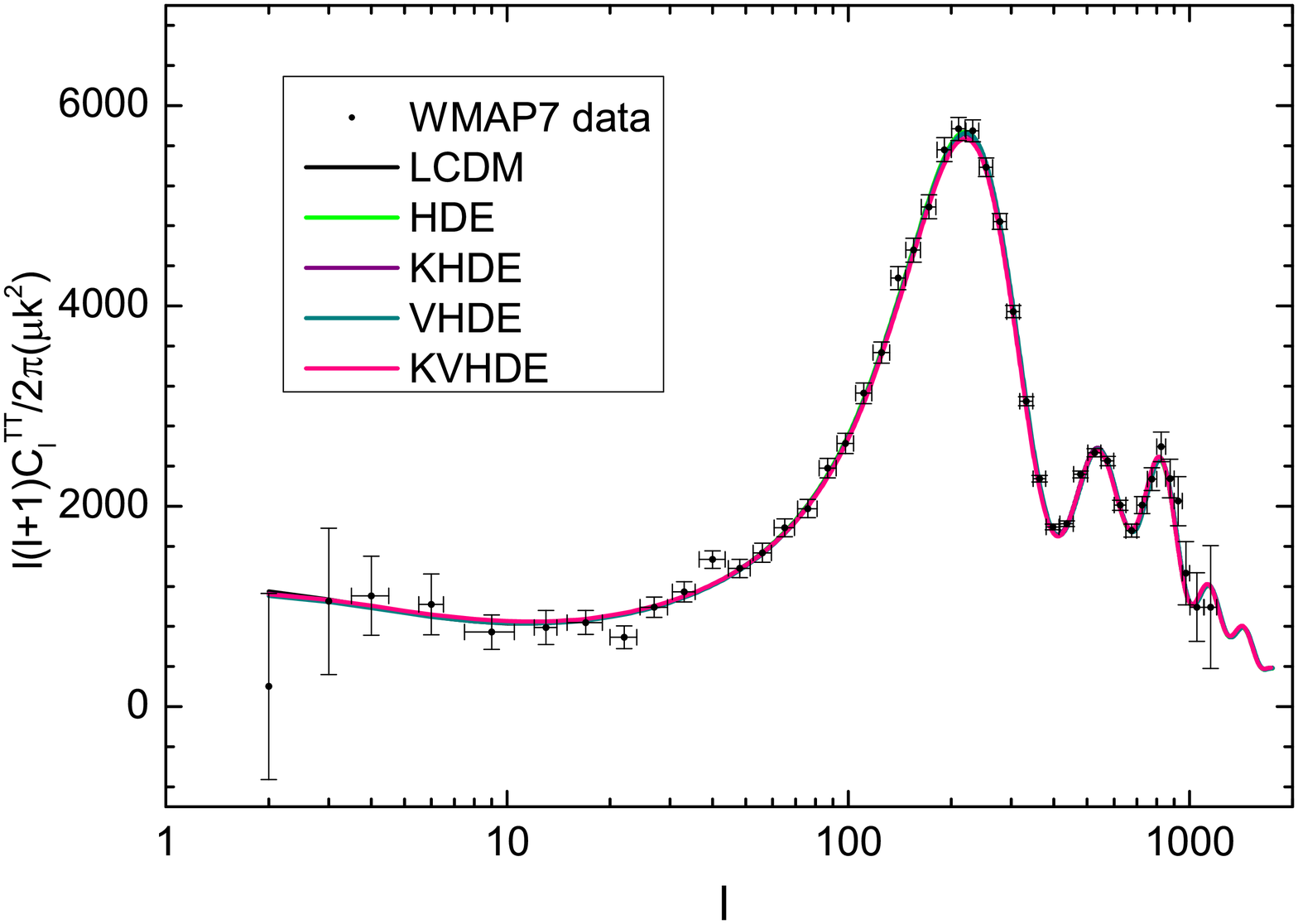}
\caption{The CMB $C_l^{TT}$ power spectra for the HDE, KHDE, VHDE
and KVHDE models with the corresponding best-fit parameter values.
The black dots with error bars denote the observed data with their
corresponding uncertainties from WMAP 7-yr results.
}\label{fig:TTpower-bf}
\end{figure}

In Fig.~\ref{fig:TTpower-bf}, we plot the CMB $C_l^{TT}$ power
spectra for the HDE, KHDE, VHDE and KVHDE models with the
corresponding best-fit parameters. To make a comparison, we also
include the $\Lambda$CDM model with the best-fit parameters given by
the same set of data. It can be seen that the $C_l^{TT}$ power
spectra for the holographic dark energy are well inside the error
bars of the observational data given by the WMAP 7-yr measurements
and match the $\Lambda$CDM model very well.

Now, let us discuss the cosmological consequence of introducing
spatial curvature and massive neutrinos in the holographic dark
energy model. We are interested in the impacts of these factors on
the EOS of dark energy and the fate of the universe. Note that, if
the spatial curvature is involved in the model, the EOS of the
holographic dark energy reads
\begin{equation}
w=-{1\over 3}-{2\over 3}\sqrt{{\Omega_{de}\over c^2}+\Omega_k},
\end{equation}
In the far future ($z\rightarrow -1$),
it is clear that $\Omega_k\rightarrow 0$ and $\Omega_{de}\rightarrow
1$, and so we still have $w|_{z\rightarrow -1}=-{1\over 3}-{2\over
3c}$. Hence, even though the spatial curvature is involved, we still
hold the conclusion that $c<1$ leads to a Big Rip future singularity
while for $c>1$ this singularity is avoided.

\begin{figure}
\centering
\includegraphics[width=8cm]{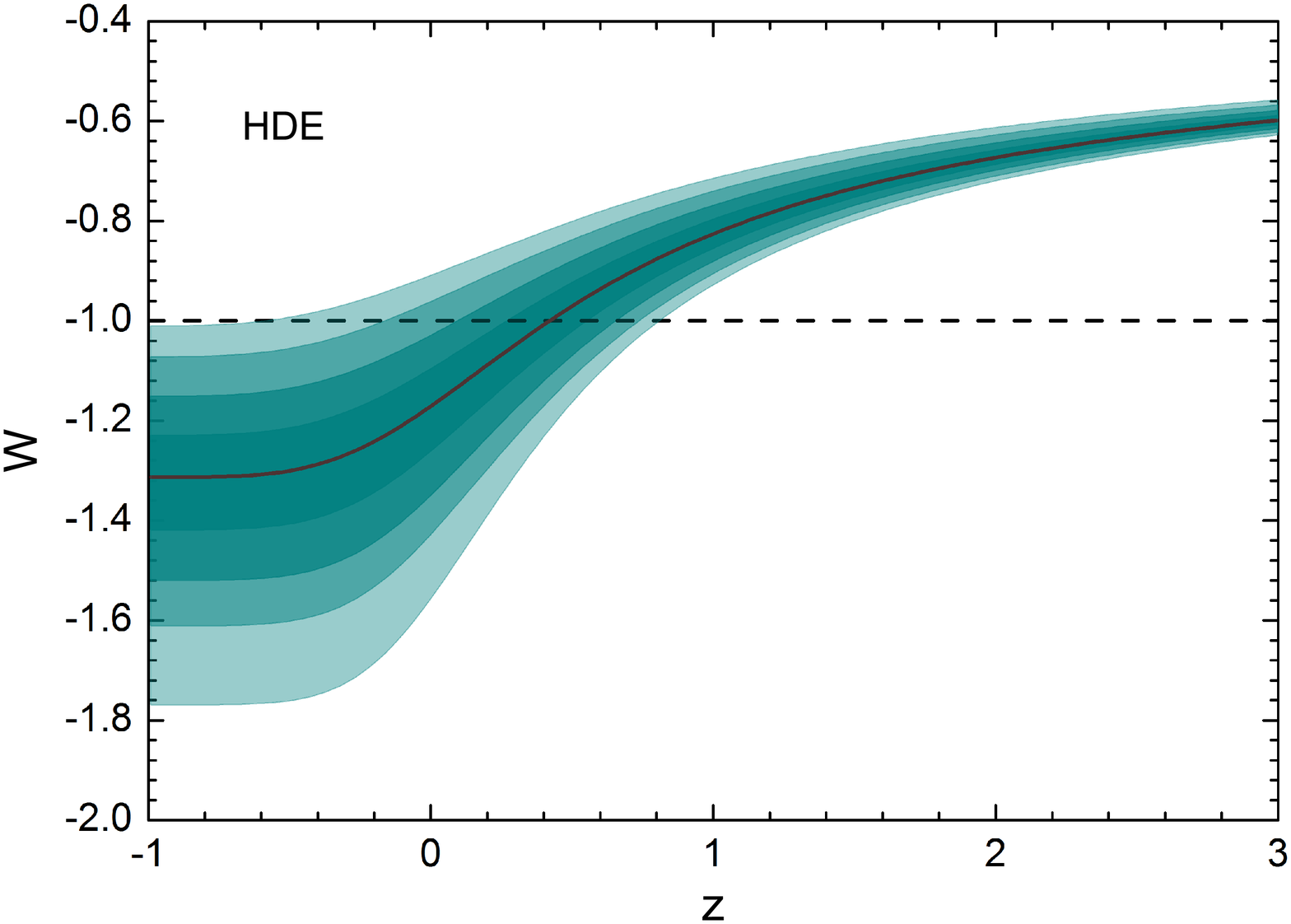}
\includegraphics[width=8cm]{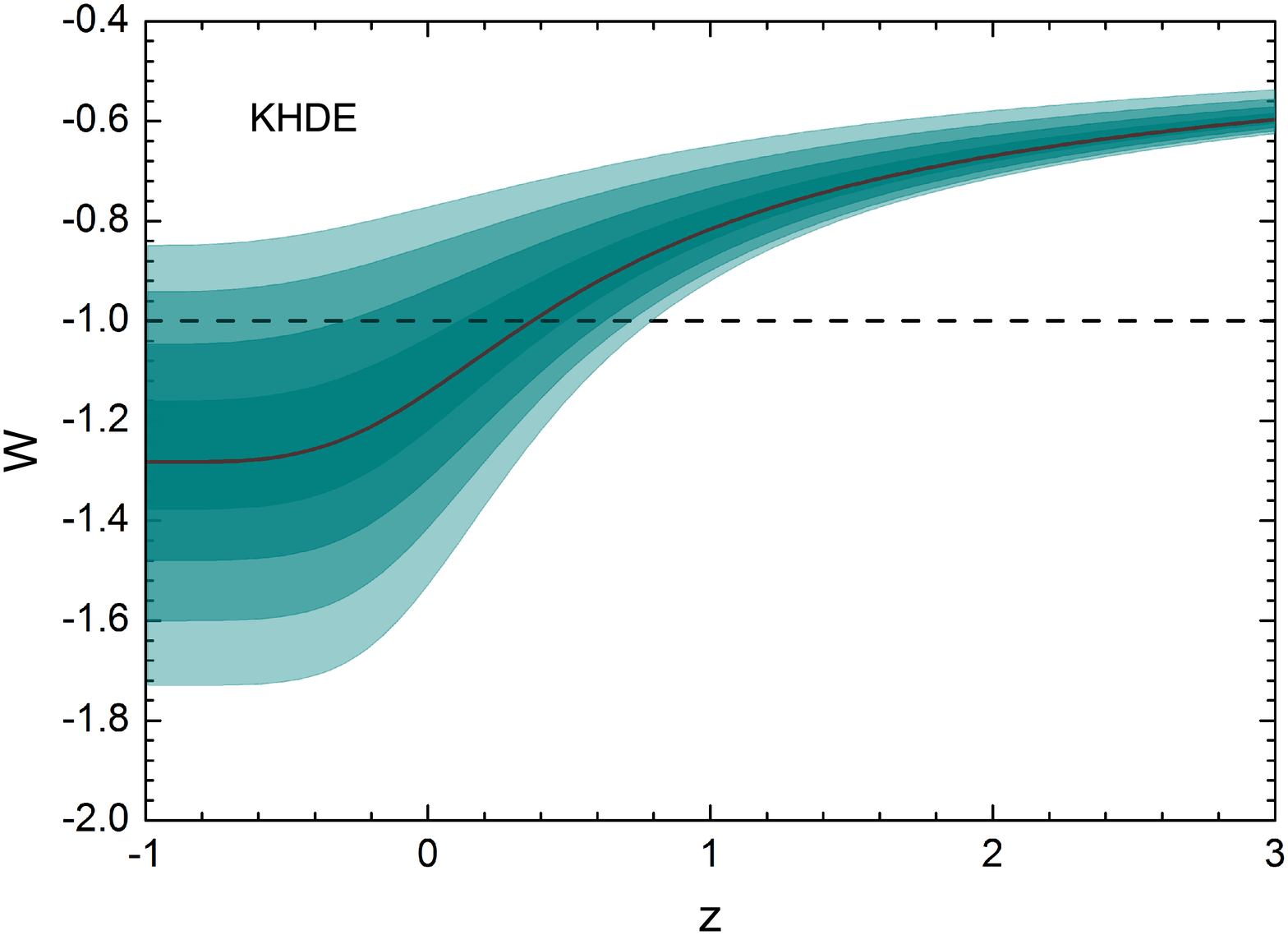}
\includegraphics[width=8cm]{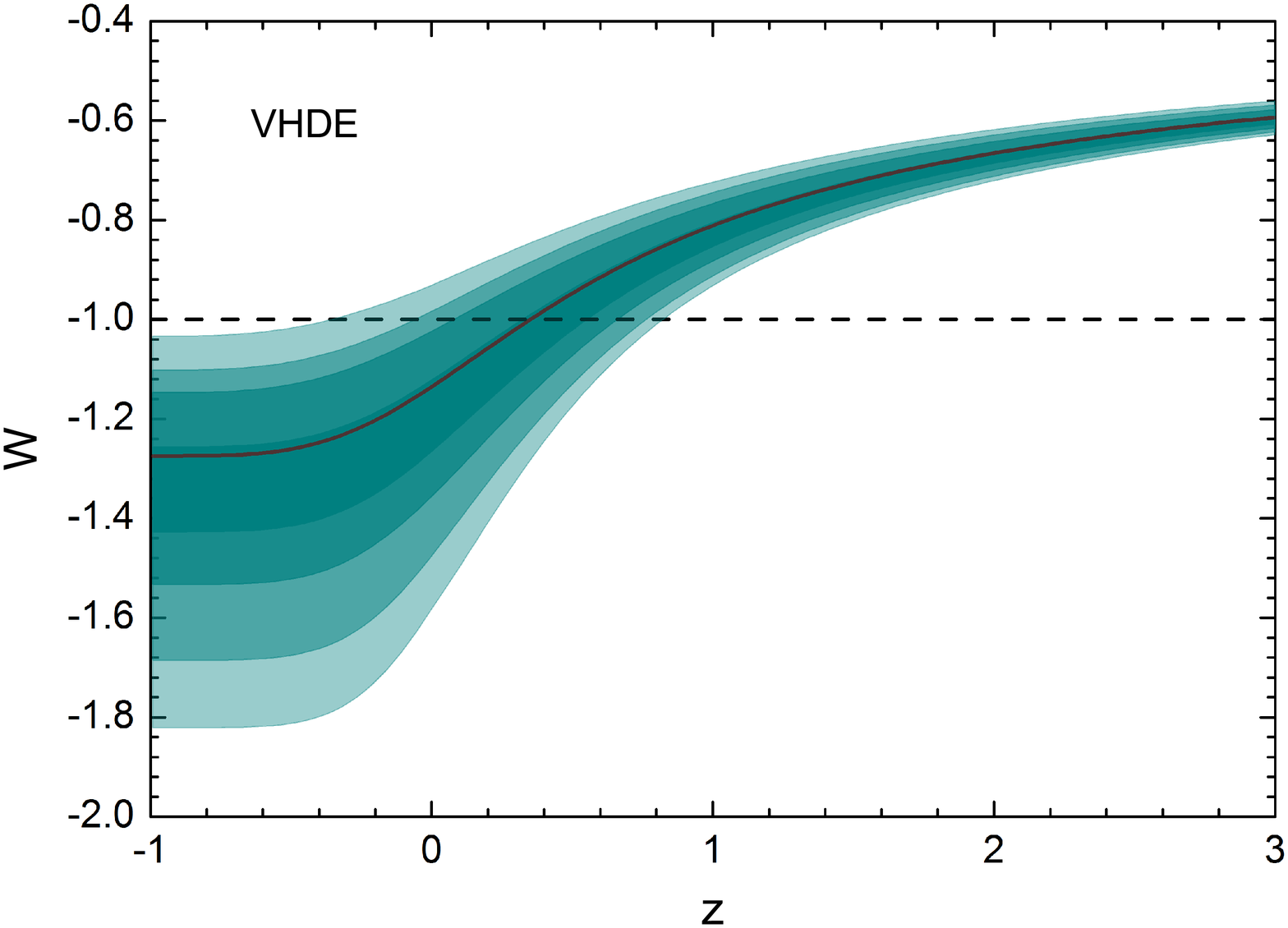}
\includegraphics[width=8cm]{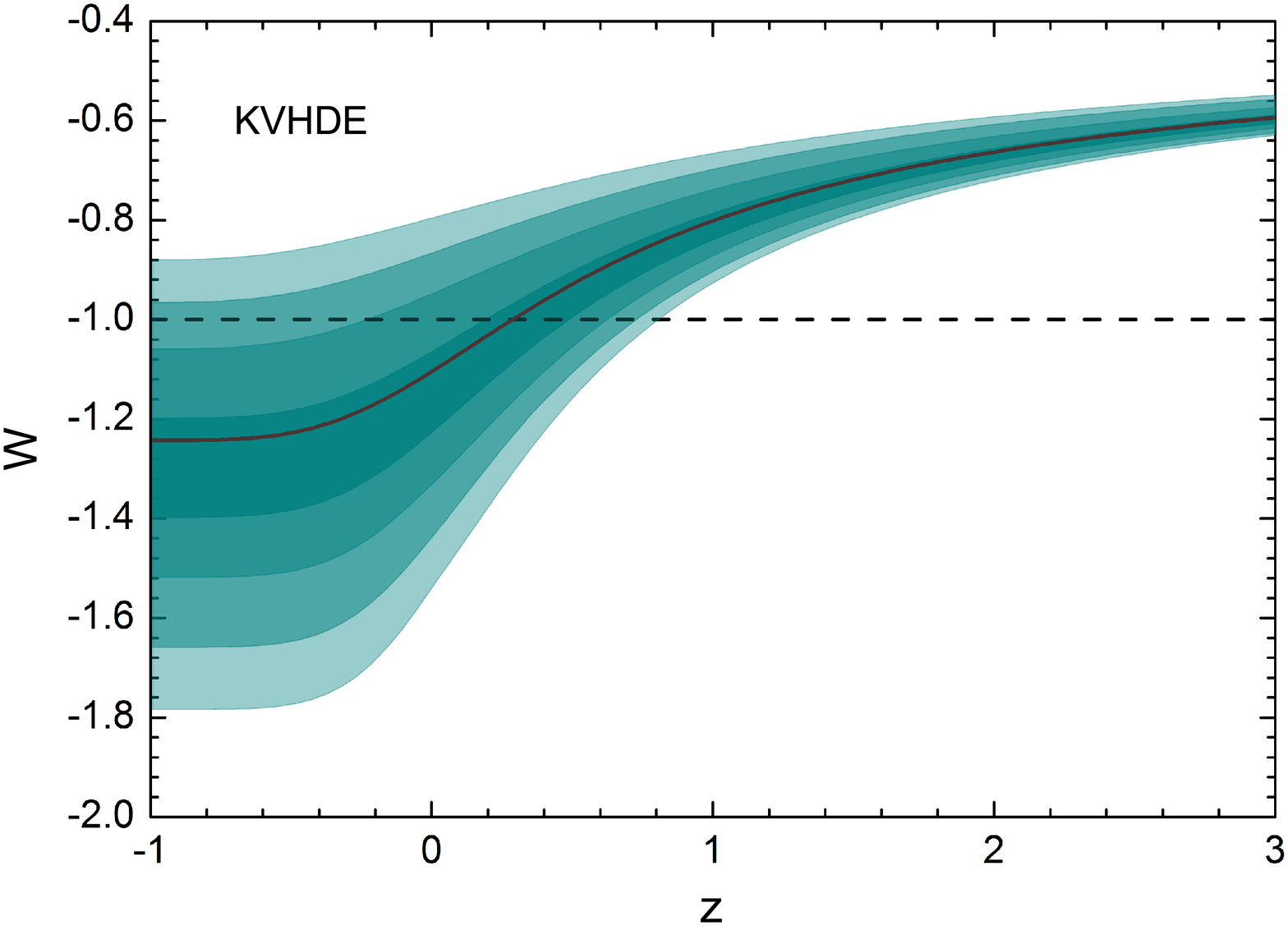}
\caption{The evolution of $w(z)$ along with redshift $z$, including
the best-fit results, as well as the 1--4$\sigma$ regions, for the
considered models.}\label{fig:wde}
\end{figure}

In Fig.~\ref{fig:wde}, we plot the evolution of $w$ along with
redshift $z$, including the best-fit results, as well as the
1--4$\sigma$ regions, for the considered models. As seen in the left
panels of Fig.~\ref{fig:wde}, for the holographic dark energy models
without spatial curvature, $w$ will cross $-1$ at more than
$4\sigma$ CL. As mentioned above, this means that the future
universe will be dominated by phantom energy and will end up with a
``Big Rip'' singularity at more than $4\sigma$ CL. As seen in the
right panels of Fig.~\ref{fig:wde}, after introducing the spatial
curvature, $w$ will cross $-1$ only in about $2.5\sigma$ range.
Therefore, the inclusion of spatial curvature in the holographic
dark energy model may be helpful to alleviate the future cosmic
doomsday problem. In contrast, the inclusion of massive neutrinos
does not have significant influence on the evolution of $w(z)$.

We also notice that the phantom behavior of the HDE model is
different from many other models. The holographic dark energy is of
phantom type at very high confidence levels. This is different from
the 2- or 3- parametric scalar field dark energy models, Chaplygin
gas models, divergence-free parametrization models, and so forth,
where the phantom type of dark energy fits the same data only
slightly better than the quintessence or cosmological constant
models
\cite{WMAP7,Wang:2012mq,Novosyadlyj:2012vd,Liao:2012gq,Li:2012via}.
This difference is due to the particular behavior of $w$ in the HDE
model. In the HDE model, there is always $w\rightarrow-\frac{1}{3}$
at high redshifts. However, from cosmological observations we know
that on average there is $w\sim -1$. Thus, to fit the data, $w<-1$
is expected at low redshifts.

\begin{figure}
\centering
\includegraphics[width=8cm]{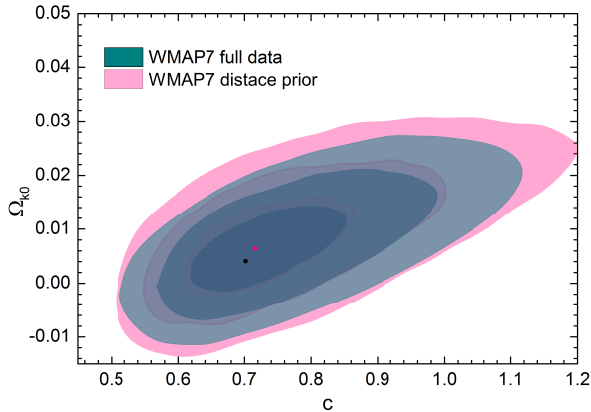}
\caption{The marginalized 1--3$\sigma$ CL contours in the
$c$-$\Omega_{k0}$ plane, for the KHDE model. The pink contours are
plotted by using the WMAP distance priors, and the olive contours
are plotted by using the full WMAP 7-yr data.}\label{fig:omnuhsq_c}
\end{figure}

We are also interested in the differences between the cosmological
constraints given by the WMAP 7-yr ``distance priors'' and those
given by the full WMAP 7-yr data. As an example, we plot the
marginalized 1--3$\sigma$ CL contours in the $c$--$\Omega_{k0}$
plane, for the KHDE model, in Fig.~\ref{fig:omnuhsq_c}. The pink
contours are plotted by using the WMAP ``distance priors'', and the
olive contours are plotted by using the full WMAP data. As seen in
this figure, the regions of the 1--3$\sigma$ contours given by the
full WMAP data are significantly smaller than the corresponding
regions given by the WMAP ``distance priors''. Therefore, making use
of the full WMAP data can give better constraints on the holographic
dark energy model, compared with the case using the WMAP ``distance
priors''.

\section{Concluding remarks}\label{sec:4}

In this paper we considered the holographic dark energy model with
spatial curvature and massive neutrinos. It is well known that both
the spatial curvature and neutrino mass are correlated with the dark
energy EOS, so it is important to study the influences of these
factors to the holographic dark energy. In addition, it is also
rather significant to consider the cosmological perturbations in
holographic dark energy and make a global fit analysis on the
holographic dark energy model.

We placed constraints on the holographic dark energy in a universe
with spatial curvature and massive neutrinos, based on a MCMC global
fit technique. The cosmic observational data include the WMAP 7-yr
temperature and polarization data, the SNIa data from Union2.1
sample, the BAO data from SDSS DR7 and WiggleZ Dark Energy Survey,
and the latest measurements of $H_0$ from HST. In order to treat the
perturbations in dark energy when $w$ crosses $-1$, we employed the
PPF approach. So, we do not suffer from the divergence problem when
$w$ crosses $-1$.

We found that, for the simplest HDE model, the phenomenological
parameter $c<1$ at more than $4\sigma$ CL, showing that the future
universe will be dominated by phantom dark energy at more than
$4\sigma$ CL. After taking into account spatial curvature, we have
$c<1$ only in about $2.5\sigma$ range, implying that the inclusion
of spatial curvature in the holographic dark energy model may be
helpful to alleviate the future doomsday problem. In contrast, the
inclusion of massive neutrinos does not have significant influence
on the phenomenological parameter $c$.

For the KHDE model, we found that the 2$\sigma$ range of the spatial
curvature is $-0.006<\Omega_{k0}<0.020$; moreover, the $3\sigma$
error bars of $\Omega_{k0}$ are still fairly small, also in order of
$10^{-2}$. For the VHDE model, we obtained the result that the
$2\sigma$ upper bound of the total mass of neutrinos is $\sum
m_{\nu} < 0.48$ eV, which is the first result of neutrino mass in
the holographic dark energy model. Moreover, when simultaneously
considering spatial curvature and massive neutrinos, the upper bound
of $\sum m_{\nu}$ will be enlarged by more than 2 times.
Furthermore, we also demonstrated that, making use of the full WMAP
7-yr data can give better constraints on the holographic dark energy
model, compared with the case using the WMAP ``distance priors''.

It should be mentioned that, there are still some factors not
covered in our paper, e.g., the possible interaction between dark
sectors. Evidently, when taking into account the interaction, the
computation of the dark sector perturbations will become much more
complicated. These issues deserve further investigations in the
future work.

\begin{acknowledgments}
We would like to thank the anonymous referee for providing us with
many helpful comments and suggestions, leading to significant
improvement of this paper. This work was supported by the National
Science Foundation of China under Grant Nos. 10705041, 10975032 and
11175042, and by the National Ministry of Education of China under
Grant Nos. NCET-09-0276 and N100505001.
\end{acknowledgments}


\end{document}